\documentclass[twocolumn]{aastex6}
\newcommand{\asap}{ A\&A\,}
\newcommand{\an}{Astron. Nachr.\,}
\newcommand{\arxiv}{ arXiv\,}
\newcommand{\basi}{ BASI\,}
\newcommand{\japa}{J. Astrophys. Astron\,}
\newcommand{\ie}{$i.e.,\;$}
\newcommand{\eg}{$e.g.,\;$}

\begin{document}


\title{J1216+0709 : A radio galaxy with three episodes of AGN jet activity}


\author{Veeresh Singh\altaffilmark{1}, C. H. Ishwara-Chandra\altaffilmark{2}, Preeti Kharb\altaffilmark{3}, Shweta Srivastava\altaffilmark{1} and P. Janardhan\altaffilmark{1}}
\altaffiltext{1}{Astronomy \& Astrophysics Division, Physical Research Laboratory, Ahmedabad 380009, India}
\altaffiltext{2}{National Centre for Radio Astrophysics, TIFR, Post Bag 3, Ganeshkhind, Pune 411007, India}
\altaffiltext{3}{Indian Institute of Astrophysics, II Block, Koramangala, Bangalore 560034, India}
%
\begin{abstract}
We report the discovery of a `Triple-Double Radio Galaxy (TDRG)' J1216+0709 detected in deep low-frequency Giant Metrewave Radio Telescope 
(GMRT) observations. 
J1216+0709 is only the third radio galaxy, after B0925+420  and Speca, with three pairs of lobes resulting from three different  episodes of AGN jet activity. 
The 610 MHz GMRT image clearly displays an inner pair of lobes, a nearly co-axial middle pair of lobes and a pair of outer lobes that is bent w.r.t. the axis of 
inner pair of lobes. 
The total end-to-end projected sizes of the inner, middle, and outer lobes are 40$^{{\prime}{\prime}}$ ($\sim$ 95 kpc), 
1$^{\prime}$.65 ($\sim$ 235 kpc) and 5$^{\prime}$.7 ($\sim$ 814 kpc), respectively. 
Unlike the outer pair of lobes both the inner and middle pairs of lobes exhibit asymmetries in arm-lengths and flux densities, 
but in opposite sense, {\ie}the eastern sides are farther and also brighter that the western sides, thus suggesting 
the possibility of jet being intrinsically asymmetric rather than due to relativistic beaming effect. 
The host galaxy is a bright elliptical (m$_{\rm r}$ $\sim$ 16.56) with M$_{\rm SMBH}$ $\sim$ 3.9 $\times$ 10$^{9}$ M$\odot$ and 
star-formation rate of $\sim$ 4.66$_{\rm -1.61}^{\rm +4.65}$ M$_{\odot}$ yr$^{-1}$. 
The host galaxy resides is a small group of three galaxies (m$_{\rm r}$ $\leq$ 17.77) and is possibly going 
through the interaction with faint, dwarf galaxies in the neighbourhood, which may have triggered the recent 
episodes of AGN activity.      

\end{abstract}


\section{Introduction} \label{sec:intro}

Radio galaxies, a subclass of Active Galactic Nuclei (AGN), are powerful radio emitters and typically exhibit radio morphology that 
consists of a core producing a pair of bipolar collimated jets terminating in the form of lobes. 
Therefore, detection of `core-jet-lobe' radio morphology is a clear indication of AGN activity. 
In fact, radio morphological structure and spectral properties can be used to probe the history of AGN activity. 
Morphological studies of radio galaxies have been useful in understanding the precession or change of jet axis in 
X-shaped radio galaxies, effect of the motion of host galaxies in bent radio galaxies, and intermittent AGN activity in radio 
galaxies showing two pairs of lobes \citep{Dehghan14,Roberts15}.
In recent years, there have been attempts to understand the details of recurrent AGN activity in galaxies 
by studying radio galaxies exhibiting two pairs of lobes that are formed during two different phases of AGN activity 
\citep[{\eg}][]{Jamrozy09,Konar13}. 
These galaxies are generally termed as `Double-Double Radio Galaxies (DDRGs)' in which a new pair of radio lobes is seen closer 
to nucleus, before distant and old pair of radio lobes fades away \citep{Saikia09}.
The new pair of lobes are edge brightened, and therefore, it can easily be distinguished from knots in the jets. \\
Interestingly, despite the identification of thousands of radio galaxies in various radio surveys only few dozens 
are confirmed DDRGs \citep{Nandi12}, possibly due to lack of sensitivity and resolution 
such that radio structures at different spatial scales remain undetected. 
Low-frequency sensitive GMRT observations with the resolution of few arcsecond are well suited to detect both steep spectrum, 
low-surface-brightness radio emission from the old pair of lobes as well as kpc-scale emission from the new pair of lobes.  
\par 
In this paper we report the discovery of rare `Triple-Double Radio Galaxy (TDRG)' named as J1216+0709 that displays three pairs of lobes in the 610 MHz GMRT image. 
This is only the third TDRG reported so far after B0925+420 \citep{Brocksopp07} and Speca \citep{Hota11}. 
This radio galaxy is either undetected or poorly detected in previous radio surveys ({\eg}Faint Images of the Radio Sky at Twenty-cm \citep[FIRST;][]{Becker95}, 
NRAO VLA Sky Survey \citep[NVSS;][]{Condon98}, VLA Low-frequency Sky Survey \citep[VLSS;][]{Cohen07}, and TIFR-GMRT Sky Survey 
\citep[TGSS;][]{Intema16} due to the lack of optimum sensitivity and resolution. 
The host galaxy has been identified in the Sloan Digital Sky Survey \citep[SDSS;][]{Ahn14} as an early type galaxy at 
RA (J2000) = 12$^{\rm h}$ 16$^{\rm m}$ 32$^{\rm s}$.42 and DEC (J2000) = +07$^{\circ}$ 09$\arcmin$ 55$\arcsec$.8 
with a spectroscopically measured redshift (z) = 0.136. \\
The cosmological parameters that we adopt are H$_{\rm 0}$ = 71 km s$^{-1}$ Mpc$^{-1}$, ${\Omega}_{\rm M}$ = 0.27 
and ${\Omega}_{\rm vac}$ = 0.73. Using this cosmology, 1 arcsec corresponds to 2.381 kpc at the luminosity distance of $\sim$ 633.8 Mpc for J1216+0709.

\section{GMRT observations and data reduction} \label{sec:observations}
The Radio galaxy J1216+0709 was observed with the GMRT at 610 MHz on 20 June 2012 and at 325 MHz on 23 April 2016. 
During our GMRT observations, we used the full array of 30 antennas and the software backend with receiver 
bandwidth of 32 MHz subdivided into 256 channels. 
During the 610 MHz observations the target field centered at NGC 4235 (RA (J2000) = 12$^{\rm h}$ 17$^{\rm m}$ 09$^{\rm s}$.9 and DEC = +07$^{\circ}$ 11$\arcmin$ 30$\arcsec$) 
was observed for nearly four hours. 
While, in 325 MHz observations the field was centered on the target source J1216+0709 and was observed for nearly 3.5 hours. 
The amplitude calibrators were observed for $\sim$ 20 minutes 
at the start and/or end of each run, and the phase calibrators were observed for $\sim$ 5 minutes in every $\sim$ 40 minutes. 
GMRT data were reduced and analysed in the standard way using the NRAO Astronomical Image Processing System (AIPS). 
Calibrated visibilities were Fourier transformed to create radio images by using the `IMAGR' task, 
and robust weighing scheme, where robust parameter was set to `0' (between uniform and natural weighing) at both frequencies. 
All the images were self-calibrated and primary beam corrected.
Our final maps have noise-rms $\sim$ 40 $\mu$Jy beam$^{-1}$ with synthesized beam-size $\sim$ 6$\arcsec$.2 $\times$ 4$\arcsec$.5 at 610 MHz, 
and $\sim$ 160 $\mu$Jy beam$^{-1}$ with synthesized beam-size $\sim$ 11$\arcsec$.4 $\times$ 8$\arcsec$.5 at 325 MHz. 
More details on the data reduction are presented in \cite{Kharb16}.\\ 
The total flux densities of different components were measured using the AIPS task `TVSTAT' that allows us to choose an area of any shape. 
Error on the flux density of a component was obtained by multiplying the average noise-rms to the total area of the component measured in 
the units of synthesized beams. Flux density of an individual component was obtained by considering the area shown by contours overlaid on to grey-scale image.

\section{Radio properties} \label{sec:radioprop}

\subsection{Radio Morphology}

J1216+0709 is best imaged at 610 MHz in GMRT observations, while it is either undetected or poorly detected in existing radio surveys. 
The 610 MHz GMRT image displays an inner pair of lobes, a nearly co-axial middle pair of lobes and a pair of outer lobes that are 
bent {\it w.r.t.} the inner pair of lobes (see figure~\ref{fig:GMRTImages}). 
The overall radio morphology of the source appears bent in a `C' shaped like structure 
and resembles somewhat to a Wide Angle Tail (WAT) radio galaxy. 
The total end-to-end projected sizes of the inner lobes, middle lobes and outer lobes are 40$\arcsec$ 
($\sim$ 95 kpc), 1$\arcmin$.65 ($\sim$ 235.7 kpc) and 5$\arcmin$.7 ($\sim$ 814 kpc), respectively (see figure~\ref{fig:GMRTImages}). 
The three distinct pairs of lobes can be interpreted as evidence for the three different episodes of AGN jet activity.
The inner and middle pair of lobes show clear edge-brightened structures which distinguish them from knots in the jets. 
In the 610 MHz image, there is relatively faint bridge-like radio emission connecting the two successive lobes, 
which tentatively indicates that the newly formed jets propagate outwards possibly through the jet-cocoon structure formed 
by the previous episode of activity rather than through intergalactic medium. 
\par
Radio flux densities of different components at different frequencies are given in Table~\ref{table:Radiofluxes}. 
Based on the 610 MHz image, Table~\ref{table:RadioSize} lists total radio sizes, ratios of flux densities and arm-lengths of 
the eastern and western side lobes and the luminosities of the three pairs of lobes. 
We note that for outer pair of lobes, both eastern and western lobes have similar flux densities and distance from the center ({\ie}arm-lengths), 
which suggests that the outer pair of lobes is lying nearly in the plane of sky. 
For both the inner and middle pairs of lobes, the eastern side is nearly 2.3 - 2.5 times stronger than the western side 
{\ie}ratios of flux densities of eastern-to-western lobe for inner and middle pairs are (R$_{\rm f,~in}$) $\sim$ 2.3 and 
(R$_{\rm f,~mid}$) $\sim$ 2.5, respectively (see table~\ref{table:RadioSize}). 
While arm-lengths of the eastern sides are larger in comparison to the western sides 
{\ie}the ratios of eastern to western sides are (R$_{\rm l,~in}$) $\sim$ 2.0 to (R$_{\rm l,~mid}$) $\sim$ 1.6, for the inner and middle pair, respectively. 
We note that the asymmetry shown by inner and middle doubles in our source is consistent with the general trend found in DDRGs in which 
the inner doubles tend to be more asymmetric in both its arm-length and its flux density ratios compared to the outer doubles \citep[see][]{Saikia06}. 
It is worth to note that for both inner and middle doubles the asymmetries in arm-lengths and flux densities are in opposite sense 
{\ie}the eastern lobes being farther and also brighter that the western lobes. 
The opposite asymmetry is difficult to explain by simple version of relativistic beaming effect. 
Also, in radio galaxies the viewing angle is relatively large ($>$ 45$^{\circ}$; \citet{Urry95}), and therefore, 
relativistic beaming effect is unlikely to be dominant. 
This strengthens the possibility of jet being intrinsically asymmetric. 
Indeed, asymmetric jet have been suggested for some of DDRGs reported in the literature \citep[see][]{Jamrozy09}. 
In our TDRG the outer pair of lobes are bent {\it w.r.t.} the axis of the inner pair of lobes and this can be understood if the 
outer lobes are entraining into a medium having large-scale density gradients or host galaxy has moved during the two cycles of AGN activity \citep{Kantharia09}.   
\\
The 610 MHz radio luminosity of the inner, middle and outer pair of lobes are 3.5 $\times$ 10$^{23}$ W Hz$^{-1}$, 5.4 $\times$ 10$^{23}$ W Hz$^{-1}$ and 
4.8 $\times$ 10$^{24}$ W Hz$^{-1}$, respectively (see table~\ref{table:RadioSize}). 
The higher luminosity of the outer pair of lobes in comparison to the inner pair of lobes 
is consistent with other cases of TDRGs and DDRGs \citep[{see}][]{Schoenmakers2000}. 
The total 610 MHz radio luminosity (5.8 $\times$ 10$^{24}$ W Hz$^{-1}$) of our TDRG 
(with r-band absolute magnitude (M$_{\rm R}$) $\sim$ -20.69$\pm$0.03 \citep{Simard11}) is close to the separation line between FR I and FR II 
types \citep{Fanaroff74}. Interestingly, the western outer lobe with edge-brightening is similar to FR II, while eastern one bears resemblance to a FR I type. 
\\
We note that there is no clear detection of the AGN core in both the 610 MHz and 325 MHz images. 
While core is marginally detected in FIRST image at 2${\sigma}$ level ($\sim$ 0.45 mJy). 
Since our 610 MHz images have typical noise-rms of 0.04 mJy and this gives an upper limit of core flux density of 0.12 mJy at 3$\sigma$ level. 
Therefore, AGN radio core exhibits inverted spectral index between 610 MHz to 1.4 GHz (see table~\ref{table:Radiofluxes}). 
The compact inverted-spectrum AGN core is similar to Giga-Hertz Peaked-spectrum Sources (GPSs) that exhibit 
peak in their radio Spectra between 1 GHz to 5 GHz and an inverted-shape spectrum at lower frequencies \citep{Fanti09}. 
GPSs are interpreted as radio AGN in early phase of their evolution, and therefore, compact inverted-spectrum radio core of TDRG 
may be considered as an indication of recent AGN activity \citep{Randall11,Orienti16}. 
Indeed, some DDRGs are known to show mildly inverted spectrum of the core \citep{Machalski10}. 

\subsection{Radio spectrum}
\begin{figure*}
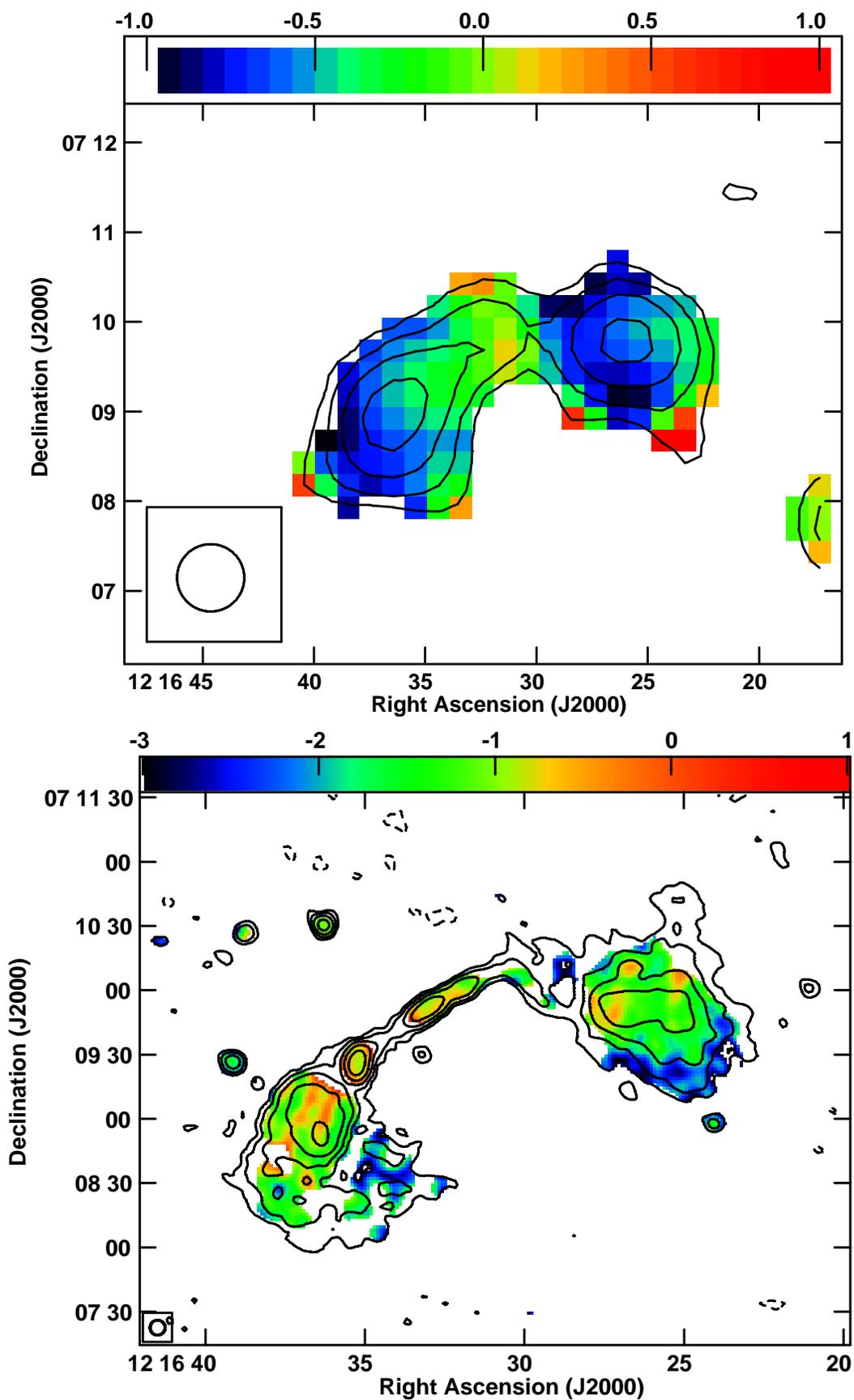

\centering
\includegraphics[angle=0,width=14.0cm,trim={0.0cm 0.0cm 0.0cm 0.0cm},clip]{fig2a.ps} 
\includegraphics[angle=0,width=14.0cm,trim={0.5cm 0.1cm 0.2cm 0.25cm},clip]{fig2b.ps}
\caption{{\it Top panel} : Spectral index map between 610 MHz (GMRT) and 1.4 GHz (NVSS) with 
NVSS contours overlaid on to it. {\it Bottom panel} : Spectral index map between 325 MHz and 610 MHz with 610 MHz contours overlaid on to it.}
\label{fig:RadioSpectrum} 
\end{figure*}

Figure~\ref{fig:RadioSpectrum} shows spectral index maps between 610 MHz and 1.4 GHz, and between 325 MHz and 610 MHz. 
We use the task `COMB' in AIPS to create spectral index images after considering flux density values above 2.5$\sigma$ at both frequencies. 
The resolutions of the images at two frequencies were matched by convolving the higher resolution image with a 
Gaussian equivalent to the beam-size of lower resolution image. 
Spectral index map between 610 MHz (GMRT) and 1.4 GHz (NVSS) shows that the outer lobes have steeper spectral index 
($\alpha$ $\sim$ - 1.0, where S$_{\rm {\nu}}$ $\propto$ ${\nu}^{\alpha}$), while the inner region 
(covering core and inner lobes) have relatively less steep spectral index ($\alpha$ $>$ - 0.5). 
325 MHz - 610 MHz spectral index map is of higher spatial resolution and shows all three lobes. 
We note that, in 325 MHz - 610 MHz spectral index map, the outer lobes have steeper spectral index ($\alpha$ $\leq$ - 1.0) in compared to the inner and middle lobes. 
Also, outer edges of the outer lobes have very steep spectral index ($\alpha$ $\leq$ - 2.0) which is typically seen in relic plasma. 
There is no signature of relatively compact hotspot-like structures with less steep spectral index, and therefore, suggesting that 
the supply of jet material has stopped long ago and hotspots in the outer lobes have completely faded away, if they existed.

\subsection{Kinematic age estimates}
It is important to estimate the time-scales of active and quiescent phase of AGN activity to understand the cause of 
episodic AGN activity and its duty cycle. Given the lack of multi-frequency radio data for our TDRG 
we can only constrain the lower limit to the active and quiescent phase time-scales by using kinematic age 
estimates based on projected radio sizes and a reasonable assumed value for the jet speed. 
We note that, in general, for a constant jet power the advancement speed of the head of lobe is higher at 
an early phase of evolution and it decreases (via interaction with the surrounding medium) as the source size increases \citep{An12b}. 
Based on the previous studies of young as well as evolved radio galaxies, we adopt the average speed of the advancement of 
the head of outer, middle and inner lobes of our source to be 0.01c, 0.05c and 0.1c, respectively, (where `c' is the speed of light) 
\citep[{see}][]{Konar06,Machalski10,An12a}. 
With these assumed speeds and the projected linear sizes of $\sim$ 814 kpc, $\sim$ 235 kpc and $\sim$ 95 kpc for the outer, middle and inner doubles, respectively, 
we obtain kinematic ages $\sim$ 1.3 $\times$ 10$^{8}$ years, $\sim$ 7.6 $\times$ 10$^{6}$ years and $\sim$ 1.5 $\times$ 10$^{6}$ years for the 
outer, middle and inner doubles, respectively.
However, we caution that our estimates of kinematic ages are only order-of-magnitude approximation under the 
simplified assumption that the advancement speed of lobe remains constant.  
We also attempt to put constraint on the time-scale of quiescent phase between the two AGN episodes. 
The duration of quiescent phase is equivalent to the time interval between the last jet material being injected in to the outer lobes 
and the first jet material being ejected from AGN in the next episode. 
The time elapsed since the last injection of relativistic particle in the jet can be estimated by synchrotron ageing method, 
which requires modelling of radio spectrum to obtain break frequency that is related to the age of synchrotron emitting plasma. 
Due to the lack of multi-frequency radio data necessary for spectral modelling we can only put a lower limit to the quiescent phase 
time-scale (t$_{\rm q}$) by subtracting the kinematic ages of outer (t$_{\rm out-lobe}$) and inner doubles (t$_{\rm in-lobe}$) 
{\ie}t$_{\rm q}$ $\leq$ t$_{\rm out-lobe}$ - t$_{\rm in-lobe}$.  
Therefore, the quiescent phase time-scales between outer (first episode) and middle (second episode) doubles, 
and the middle (second episode) and inner (third episode) are, $\leq$ 1.2 $\times$ 10$^{8}$ years and $\leq$ 6.1 $\times$ 10$^{6}$ years, respectively. 
These limits on the quiescent phase time-scales are consistent with the time-scale estimates in DDRGs \citep[{see}][]{Konar13}.  
Our proposed multi-frequency GMRT and VLA observations will allow us to put much stronger constraints on the time-scales via spectral aging measurements.

\begin{figure*}
\includegraphics[angle=0,width=18.0cm,trim={2.25cm 2.0cm 2.5cm 2.25cm},clip]{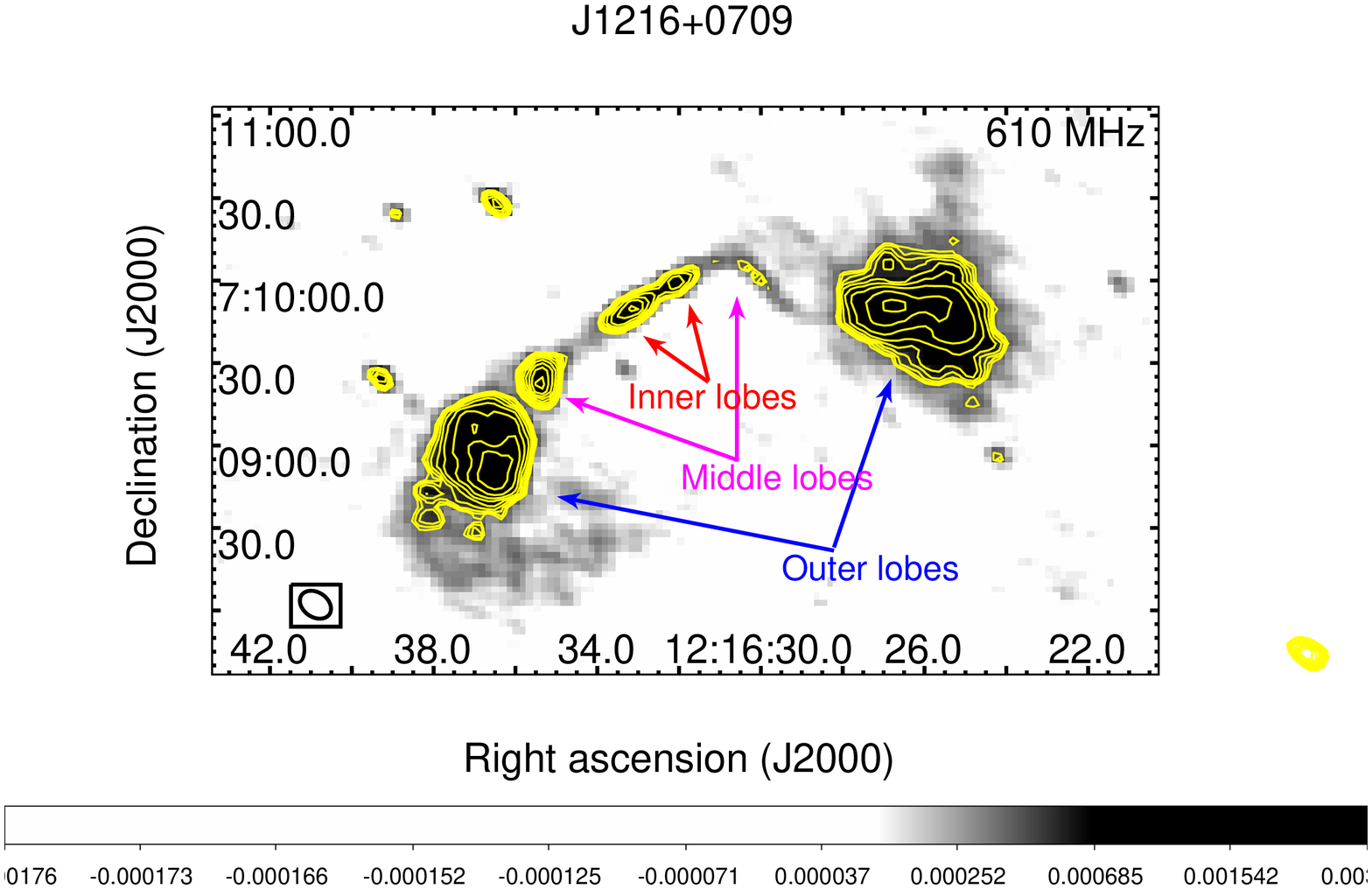}
{\includegraphics[angle=0,width=9.0cm,trim={1.5cm 1.9cm 2.75cm 2.0cm},clip]{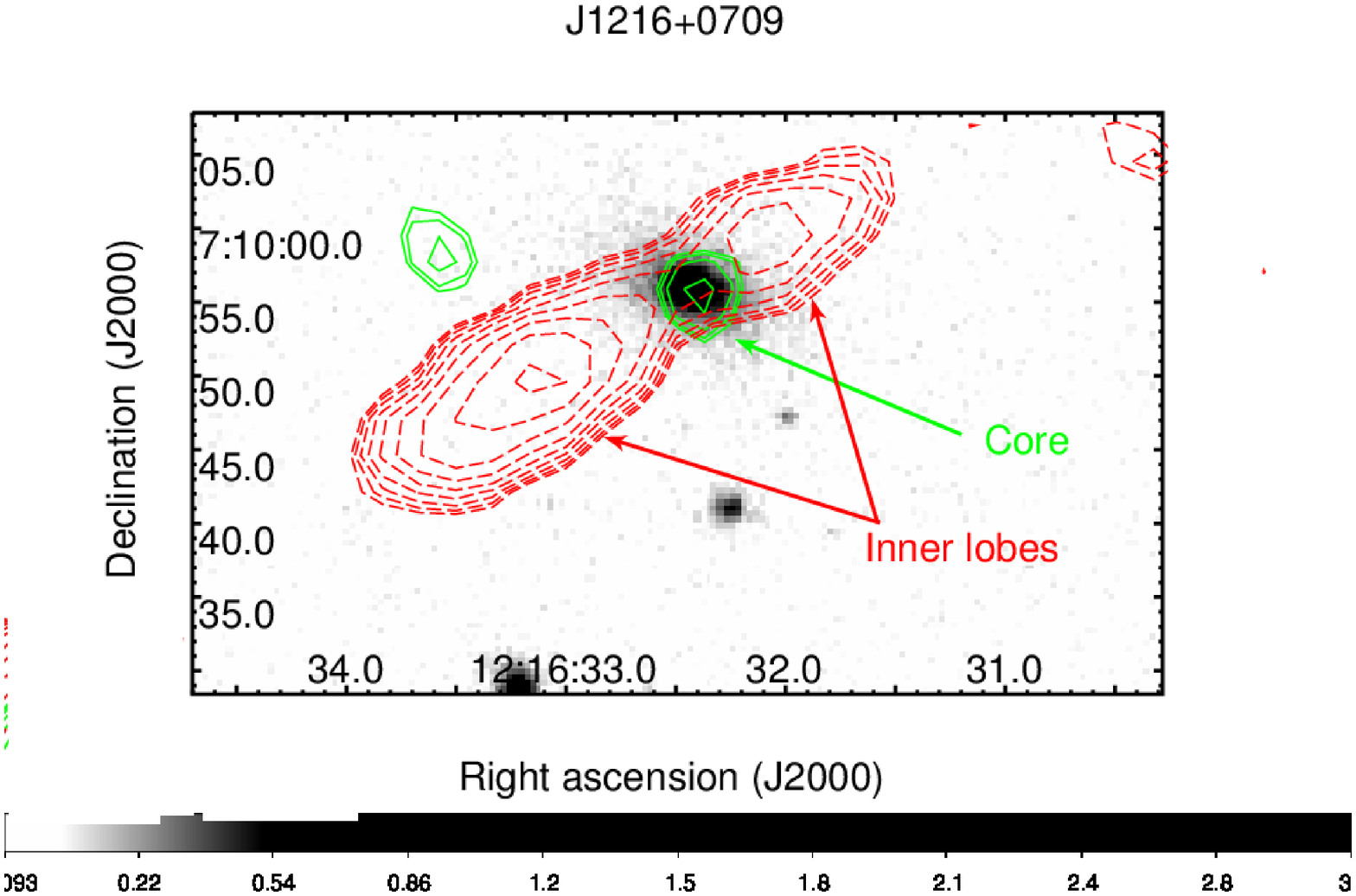}}{\includegraphics[angle=0,width=9.0cm,trim={1.5cm 1.9cm 2.75cm 2.0cm},clip]{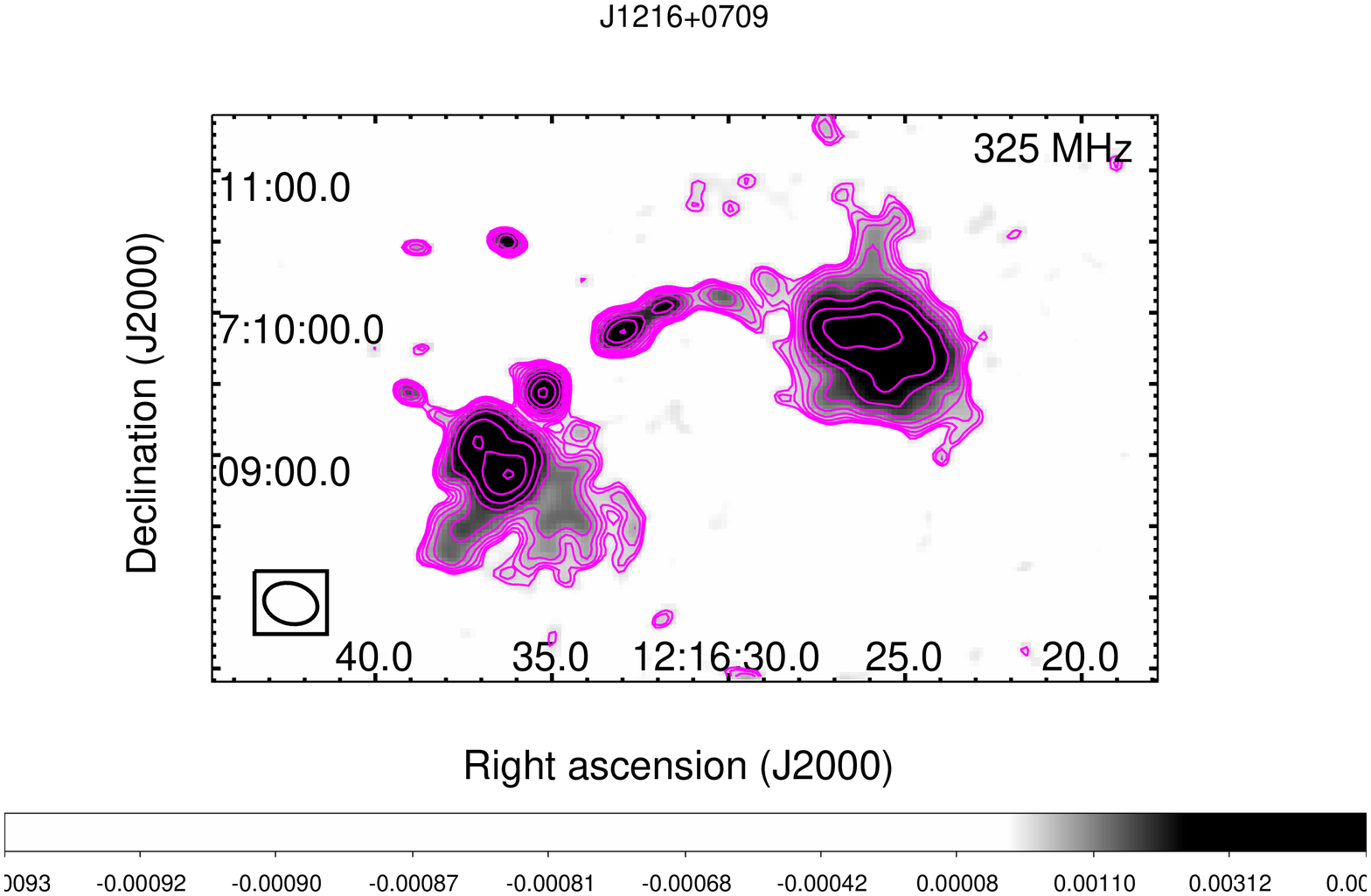}}
\caption{{\it Top panel}: The 610 MHz image displaying three pairs of lobes. 
For better visualization of different lobes the contours represent only bright emission with the lowest contour at $>$ 6$\sigma$ level.
{\it Bottom left panel}: Zoom in view of the 610 MHz contours of the inner lobes and 1.4 GHz FIRST$-$detected core (in Green) overlaid on the SDSS r-band image. 
{\it Bottom right panel}: 325 MHz GMRT image with contours overlaid on to the Grey-scale. Contours are spaced in logarithmic scale with 
the lowest contour at 5$\sigma$ level. The synthesized elliptical beam enclosed in a rectangular box is shown at the bottom left corner.}
\label{fig:GMRTImages} 
\end{figure*}

\begin{table*}
\centering
\begin{minipage}{140mm}
\caption{Radio properties}
\begin{tabular}{@{}cccccc@{}}
\hline
                              & S$_{\rm int,~325~MHz}$ & S$_{\rm int,~610~MHz}$ & S$_{\rm int,~1.4~GHz}$ & ${\alpha}_{\rm 325~MHz}^{\rm 610~MHz}$ & ${\alpha}_{\rm 610~MHz}^{\rm 1.4~GHz}$ \\  
                              &  (mJy)            &  (mJy)            &  (mJy)       &                                &                     \\ \hline
Inner lobe (E)                & 8.2$\pm$0.8       & 5.2$\pm$0.5       &              &  -0.72$\pm$0.06    &                     \\
Inner lobe (W)                & 3.8$\pm$0.6       & 2.3$\pm$0.4       &              &  -0.80$\pm$0.10    &                     \\
Middle lobe (E)               & 11.9$\pm$1.2      & 8.4$\pm$0.6       &              &  -0.55$\pm$0.05    &                     \\
Middle lobe (W)               & 5.4$\pm$1.5       & 3.7$\pm$1.8       &              &  -0.60$\pm$0.24    &                     \\
Outer lobe (E)                & 75$\pm$11.5       & 47.9$\pm$13.8     & 20.0$\pm$1.4 &  -0.71$\pm$0.14    &  -1.05$\pm$0.13      \\
Outer lobe (W)                & 100$\pm$12        & 55.8$\pm$19.6     & 21.3$\pm$1.5 &  -0.92$\pm$0.16    &  -1.16$\pm$0.16      \\
 Total                        & 207$\pm$32        & 125$\pm$14.6      & 51.5$\pm$2.4 &  -0.80$\pm$0.08    &  -1.07$\pm$0.05      \\
 Core                         & $<$ 0.48$\pm$0.16 & $<$ 0.12$\pm$0.04 & 0.45$\pm$0.2 &                          & $>$ +1.59$\pm$0.24  \\
\hline
\end{tabular}
\label{table:Radiofluxes} 
\\
Note - Outer lobe flux densities at 1.4 GHz are only crude estimates based on the area covered by outer lobe seen in the 610 MHz image. 
There is no clear detection of AGN core in 325 MHz and 610 MHz image and 
hence we put 3$\sigma$ lower limit at these frequencies.  
\end{minipage}
\end{table*}

\begin{table*}
\begin{minipage}{140mm}
\caption{Radio sizes and luminosities}
\begin{tabular}{@{}ccccccccccccc@{}}
\hline
    l$_{\rm in}$  &  l$_{\rm mid}$  &  l$_{\rm out}$  & R$_{\rm l,~in}$ & R$_{\rm l,~mid}$ & R$_{\rm l,~out}$ &  R$_{\rm f,~in}$ & R$_{\rm f,~mid}$ & R$_{\rm f,~out}$ & P$_{\rm in,~610~MHz}$  & P$_{\rm mid,~610~MHz}$  & P$_{\rm out,~610~MHz}$ & P$_{\rm total,~610~MHz}$ \\
      (kpc)       &    (kpc)        &   (kpc)         &                 &                  &                  &                  &                  &                  &  (W Hz$^{-1}$)         &  (W Hz$^{-1}$)          &       (W Hz$^{-1}$)    &       (W Hz$^{-1}$)   \\ \hline
       95         &    235          &    814          &  2.0            &   1.6            &  1.0             &     2.28         &  2.5             &    0.85          & 3.5 $\times$ 10$^{23}$ & 5.4 $\times$ 10$^{23}$  & 4.8 $\times$ 10$^{24}$ & 5.8 $\times$ 10$^{24}$ \\
\hline
\end{tabular}
\label{table:RadioSize} 
\\
Note - l$_{\rm in}$, l$_{\rm mid}$ and l$_{\rm out}$ represent total end-to-end linear sizes of inner, middle and outer pairs of lobes, respectively. 
R$_{\rm l,~in}$, R$_{\rm l,~mid}$ and R$_{\rm l,~out}$ are ratios of the linear sizes of eastern to western sides of the inner, middle and outer pairs of lobes, respectively. 
R$_{\rm f,~in}$, R$_{\rm f,~mid}$ and R$_{\rm f,~out}$ are 610 MHz flux density ratios of eastern to western lobes for the inner, middle and outer pairs of lobes, respectively.  
P$_{\rm in,~610~MHz}$, P$_{\rm mid,~610~MHz}$ and P$_{\rm out,~610~MHz}$ are 610 MHz luminosity of inner, middle and outer pair of lobes, respectively. 
The radio luminosities are k-corrected by assuming that the radio emission is synchrotron emission characterized by a power law (S$\nu$ $\propto$ ${\nu}^{\alpha}$), 
where we use the spectral index value measured between 325 MHz and 610 MHz. 
The radio luminosity of the source at redshift z and luminosity-distance d$_{\rm L}$ is, therefore, given 
by P$\nu$ =  4$\pi$d$_{\rm L}^{2}$ S$\nu$ (1 + z)$^{\rm -({\alpha}+1)}$.

\end{minipage}
\end{table*}

\section{Host galaxy and its large-scale environment}
Host galaxy and its surrounding large-scale environment can play crucial role in triggering episodic AGN activity, therefore, 
we examine the nature of host galaxy and its environment. 
Radio contours overlaid on the SDSS optical image identify the host galaxy as an elliptical galaxy at the redshift of $\sim$ 0.136 
(see figure~\ref{fig:GMRTImages}). 
Host galaxy is fairly bright (r-band magnitude $\simeq$ 16.56) and redder in colour (u - r = 2.89). 
The SDSS optical spectrum is dominated by a red continuum with noticeable 4000 {\AA} break, a characteristic of early-type galaxies. 
The optical emission line flux ratio diagnostic classifies the host galaxy as a Low Excitation Radio Galaxy (LERG). 
The Wide-field Infrared Survey Explorer \citep[WISE;][]{Wright10} colors of the host galaxy ([3.4] - [4.6] = 0.12 $<$ 0.8 in Vega magnitudes) indicate that the mid-IR emission is dominated by star formation 
and AGN contamination is not significant \citep[see][]{Stern12}.
We estimate star formation rate (SFR) of $\sim$ 4.66$_{\rm -1.61}^{\rm +4.65}$ M$_{\odot}$ yr$^{-1}$ in the host galaxy using \citet{Kennicutt98} empirical relation 
(SFR (M$_{\odot}$ yr$^{-1}$) = 4.5 $\times$ 10$^{-44}$ L$_{\rm 8~-~1000~{\mu}m}$ (erg s$^{-1}$)) based on IR luminosity. 
The total IR (8 - 1000 $\mu$m) luminosity is estimated from the WISE 22 $\mu$m luminosity using the full range of templates in the 
libraries of \citet{Chary01} and \citet{Dale02}. 
For our TDRG we estimate the mass of the super-massive black hole (M$_{\rm SMBH}$) to be $\sim$ 3.87 $\times$ 10$^{9}$ M$\odot$ 
by using the `black hole mass -bulge luminosity' relation for early type galaxies given in \citet{McConnell13}. 
The absolute bulge magnitude of our source is taken from \citet{Simard11} who present bulge$-$disc decomposition for SDSS DR7 galaxies. 
To examine if our TDRG is associated with any cluster or group we use the catalogue of \citet{Tempel14}
who identified galaxy groups and clusters based on a modified friends-of-friends method and
present a flux (m$_{\rm r}$ $\leq$ 17.77) and volume-limited catalogue using SDSS DR 10 data. 
The value of flux limit is based on the fact that the SDSS data are incomplete for fainter sources \citep{Strauss02}.
According to \citet{Tempel14} catalog, the host galaxy is part of a small group of three galaxies 
(other two galaxies located at RA (J2000) = 12$^{\rm h}$ 16$^{\rm m}$ 39$^{\rm s}$, DEC (J2000) = +07$^{\circ}$ 10${\arcmin}$ 25${\arcsec}$, 
and RA (J2000) = 12$^{\rm h}$ 16$^{\rm m}$ 30$^{\rm s}$, DEC (J2000) = +07$^{\circ}$ 06${\arcmin}$ 14${\arcsec}$) with a total estimated mass 
of $\sim$ 1.99 $\times$ 10$^{13}$ M$\odot$ and virial radius of $\sim$ 0.28 Mpc. 
\\
There is no apparent disturbance in the host galaxy morphology, and thus, we can rule out any recent major merger. 
However, a minor merger or a strong interaction with a dwarf galaxy is plausible without resulting any prominent disturbance. 
In the SDSS image, two faint, blue-color, dwarf galaxy like objects 
(one at RA (J2000) = 12$^{\rm h}$ 16$^{\rm m}$ 31$^{\rm s}$.97, DEC (J2000) = +07$^{\circ}$ 09${\arcmin}$ 47${\arcsec}$.28 with m$_{\rm r}$ $\sim$ 23.31$\pm$0.27 but without any estimate of redshift due to faintness, 
and second one at RA (J2000) = 12$^{\rm h}$ 16$^{\rm m}$ 32$^{\rm s}$.25, DEC (J2000) = +07$^{\circ}$ 09${\arcmin}$ 40${\arcsec}$.92 with m$_{\rm r}$ $\sim$ 19.67$\pm$0.02 and z$_{\rm phot}$ $\sim$ 0.118$\pm$0.0402) are seen close 
to the southern side of the TDRG host galaxy. 
The apparent fuzziness and blue color may be an indication of strong interaction that might have triggered a recent star-formation. 
So, it is possible that both these dwarf galaxies are interacting with TDRG host galaxy which can bring sufficient matter close to the SMBH to trigger AGN activity. 
However, more sensitive optical observations are required to obtain spectroscopic redshifts of these dwarf galaxies, and to confirm this possibility. 
\section{Summary}
We report the discovery of a rare `Triple-Double Radio Galaxy (TDRG)' J1216+0709 that exhibits three distinct pairs of lobes 
in the 610 MHz GMRT image. This TDRG is only the third such source reported after B0925+420, and Speca, where 
three pairs of lobes are result of three different episodes of AGN jet activity.
The 610 MHz GMRT image exhibits an inner pair of lobes, a nearly co-axial middle pair of lobes and 
a pair of outer lobes that are bent w.r.t. the inner pair of lobes. The total end-to-end projected 
sizes of the inner double, middle double and outer double are 40$\arcsec$ ($\sim$ 95 kpc), 1$\arcmin$.65 ($\sim$ 235.7 kpc) 
and 5$\arcmin$.7 ($\sim$ 814 kpc), respectively. 
We note that unlike the outer pair of lobes both the inner and middle doubles exhibit asymmetries in arm-lengths and flux densities 
but in opposite sense {\ie}eastern sides are farther and also brighter that the western sides. 
The opposite asymmetry is difficult to explain by a simple version of relativistic beaming effect and suggests 
the possibility of jet being intrinsically asymmetric.
Also, all three pairs of lobes bear edge-brightened resemblance with FR II type radio galaxies, while their 
total radio luminosities are lower than that for classical FR II radio galaxies. 
Spectral index map between 325 MHz and 610 MHz shows that the outer lobes exhibit steeper 
spectral index ($\alpha$ $\leq$ -1) in compared to the middle and inner lobes. 
The lack of hotspots and very steep spectral index in the outer edges of outer lobes indicate the presence of relic plasma. 
\par
Kinematic age estimates based on assumed advancement speed of the head of the lobes to be 0.01c, 0.05c and
0.1c, for the outer, middle and inner doubles, respectively, are $\sim$ 1.3 $\times$ 10$^{8}$ years, 
7.6 $\times$ 10$^{6}$ years, and 1.5 $\times$ 10$^{6}$ years for the outer, middle and inner pair
of lobes, respectively. 
The kinematic age estimates allow us to put a lower limit on the quiescent phase time-scales between outer (first episode) and middle (second episode) doubles, 
and the middle (second episode) and inner (third episode) doubles to be $\leq$ 1.2 $\times$ 10$^{8}$ years and $\leq$ 6.1 $\times$ 10$^{6}$ years, respectively. 
The host galaxy is found to be a bright elliptical (r-band magnitude $\sim$ 16.56) for which the optical spectrum is dominated by a red
continuum and emission line ratios suggest AGN emission to be of low excitation. 
The host galaxy contains SMBH with the mass of $\sim$ 3.87 $\times$ 10$^{9}$  and exhibit SFR $\sim$ 4.66$_{\rm -1.61}^{\rm +4.65}$ M$_{\odot}$ yr$^{-1}$.
Also, host galaxy belongs to only a small group of three galaxies with the total estimated mass of the group to be $\sim$ 1.99 $\times$ 
10$^{13}$ M$\odot$ and virial radius of $\sim$ 0.28 Mpc.
There is no apparent disturbance in the morphology of host galaxy, however, it may be interacting with two nearby dwarf galaxies. 
So, the AGN activity might have been triggered by the interaction with neighboring dwarf galaxies. 
Although, more sensitive optical data are required to confirm this plausibility.

\acknowledgments
The GMRT is a national facility operated by the National Centre for Radio Astrophysics of the Tata Institute
of Fundamental Research. We thank the staff at NCRA and GMRT for their support. 
This research has made use of the NASA/IPAC Extragalactic Database (NED) which is operated by the Jet Propulsion 
Laboratory, California Institute of Technology, under contract with the National Aeronautics and Space Administration. This publication 
makes use of data products from the WISE, which is a joint project of the University of California, Los Angeles, and the Jet 
Propulsion Laboratory/California Institute of Technology, funded by the National Aeronautics and Space Administration. 
Funding for SDSS-III has been provided by the Alfred P. Sloan Foundation, the Participating Institutions, the National Science 
Foundation, and the US Department of Energy Office of Science. The SDSS-III web site is http://www.sdss3.org/. SDSS-III is managed 
by the Astrophysical Research Consortium for the Participating Institutions of the SDSS-III Collaboration including the University 
of Arizona, the Brazilian Participation Group, Brookhaven National Laboratory, Carnegie Mellon University, University of 
Florida, the French Participation Group, the German Participation Group, Harvard University, the Instituto de Astrofisica de Canarias,
the Michigan State/Notre Dame/JINA Participation Group, Johns Hopkins University, Lawrence Berkeley National Laboratory, Max
Planck Institute for Astrophysics, Max Planck Institute for Extraterrestrial Physics, New Mexico State University, New York University,
Ohio State University, Pennsylvania State University, University of Portsmouth, Princeton University, the Spanish Participation
Group, University of Tokyo, University of Utah, Vanderbilt University, University of Virginia, University of Washington, and Yale
University.

\end{document}